\documentclass{aa}
\usepackage{graphicx}

\begin{document}

\thesaurus{Section:06, 08.05.3, 08.08.1} %

\title{\bf The puzzling theoretical predictions for the luminosity of clumping He burning stars}

\author{ V. Castellani \inst{1}$^,$\inst{2},  S. Degl'Innocenti
\inst{1}$^,$\inst{2}, L. Girardi \inst{3}, M. Marconi \inst{4},
 P.G. Prada Moroni \inst{1}, A. Weiss \inst{5}}

\offprints {S. Degl'Innocenti, scilla@astr18pi.difi.unipi.it}

\institute{
Dipartimento di Fisica, Universit\'a di Pisa, piazza Torricelli 2,
   I-56126 Pisa, Italy 
\and Istituto Nazionale di Fisica Nucleare, Sezione di Pisa, via Livornese
1291, I-56010 S. Piero a Grado, Pisa, Italy
\and Dipartimento di Astronomia, Vicolo dell'Osservatorio 5, I-35122 Padova,
Italy
\and Osservatorio Astronomico di Capodimonte, via Moiariello 16, I-80131
Napoli, Italy
\and Max Planck Institut f\"ur Astrophysics, K. Schwarzschild Strasse 1,
D-85740 Garching, M\"unchen, Germany}

\date{Received ; accepted }

\authorrunning{Castellani et al.} 
\titlerunning{Theoretical scenario for He burning stars}

\maketitle


\begin{abstract}

This paper deals with theoretical predictions for He burning models 
in a range of masses covering the so-called Red Giant Branch
phase transition. Taking as a guideline the observational 
constraints given by Hipparcos parallaxes to the predicted 
luminosity  of models originated from Red Giant progenitors with 
He core undergoing  electron degeneracy, we compare models by various
authors as recently appeared in the literature, disclosing  sensitive
differences in the predicted luminosity. The "solidity" of these
theoretical predictions is investigated by exploring the effects of varying
the assumptions about the efficiency of core overshooting 
or the amount of mass loss, giving quantitative estimates of
the related uncertainties. However, one finds that theoretical 
predictions concerning the luminosity of the red giant clump
in the Hipparcos sample is scarcely affected by these mechanisms.
{\bf A comparison among theoretical predictions
as recently given by different authors convincingly demonstrates
that the different luminosity predictions are the natural results of evolutionary codes
with different - but all reasonable- input physics.}  
In this context observations suggests that stellar models based on the "most updated" input 
physics are possibly overestimating the luminosity of these structures, 
raising doubts on several current predictions concerning  the luminosity 
of HB stars in galactic globulars.

\end{abstract}

\keywords{Stars:evolution, Stars: Hertzsprung-Russell diagram}


\section{Introduction}

Hipparcos parallaxes for clumping He burning Red Giants in the 
solar neighborhood have recently raised a large interest in
the astronomical community, providing a new valuable (though controversial)
tool to approach the problem of Magellanic Clouds  distances (see, 
e.g., Udalski et al. 1998, Stanek, Zaristski et al. 1998, Cole 
1998, Girardi et al. 1998). However, on theoretical grounds  one has to
notice
that these parallaxes provide us for the first time with direct
observational
evidences for the luminosity of He burning stars whose Red Giant
progenitors
experienced  electron degeneracy in the stellar core. Therefore providing  
a relevant test for the rather sophisticated input physics supporting 
current theoretical predictions concerning galactic globulars and, more in 
particular, concerning some  relevant issues
as the luminosity of RR Lyrae stars and the ages of halo stellar clusters.

Such a test appears now of particular interest because of the
growing rumor  about a possible overluminosity
of theoretical models for He burning stars with degenerated progenitors. 
As a matter of example, 
Pols et al. (1998) found  that the He clump luminosity in the open 
cluster M67 appears 0.2 - 0.3 mag fainter than predicted on the basis 
of their evolutionary computations. Such a discrepancy appears
larger than the expected uncertainties on the bolometric correction
and it appears further supported by independent evolutionary 
computations recently presented by Castellani et al. (1999),
hereinafter C99, which 
have discussed the severe difficulties in fitting the  
CM diagram of that clusters. However, in the meantime, Girardi et al. 
(1998)  presented a careful discussion on the luminosities of 
Hipparcos He burning giants,  which were found in
splendid agreement with  the adopted  theoretical predictions. 

To throw light on such a scenario, in the next section
we will discuss current predictions about the luminosity of He burning
stars in a suitable mass range, disclosing the occurrence of 
sensitive differences among various authors. Section 3 will be devoted to a
preliminary investigation of the uncertainties in theoretical results due
to uncertainties  on both the amount of extramixing from convective
cores and the amount of mass loss in the pre-He burning phases.
On this basis, in section 4 we will compare theoretical results 
with the observed  mean  luminosity of the red giant clump in the 
Hipparcos sample, concluding for the possible need of a revision of the
"most updated" input physics used  in recent evolutionary models.
The origin of differences in the predicted luminosities are 
finally discussed in section 5, by comparing the results of 
selected evolutionary codes to the light of the adopted physical ingredients.
A short section of concluding remarks will close the paper.

\section {Comparing theories}

According to  Girardi et al. (1998), the Hipparcos sample of neighboring
He burning giants is largely populated by stars with masses
below or in the range of the so-called Red Giant Branch phase transition
(RGB-pt). Since the pioneering paper by Sweigart et  al. (1990) this
range of stellar masses has been the subject of several careful 
evolutionary investigations. As well known, as we go from stars with 
M$\sim$1 M$_{\odot}$  to higher masses, we progressively find 
H-shell burning stars with a lower degree of electron degeneracy 
in their cores. Eventually,  He is quietly ignited 
in the center of the structure. As a consequence, the mass of the He core
at the He ignition progressively decreases, reaching a minimum
at a star mass  which depends  on the original chemical composition.
After this minimum, the He core grows again with mass, following 
the increasing size of the central convective core in the Main Sequence 
structures.

\begin{figure}[h]
\includegraphics[scale=0.80,draft=false]{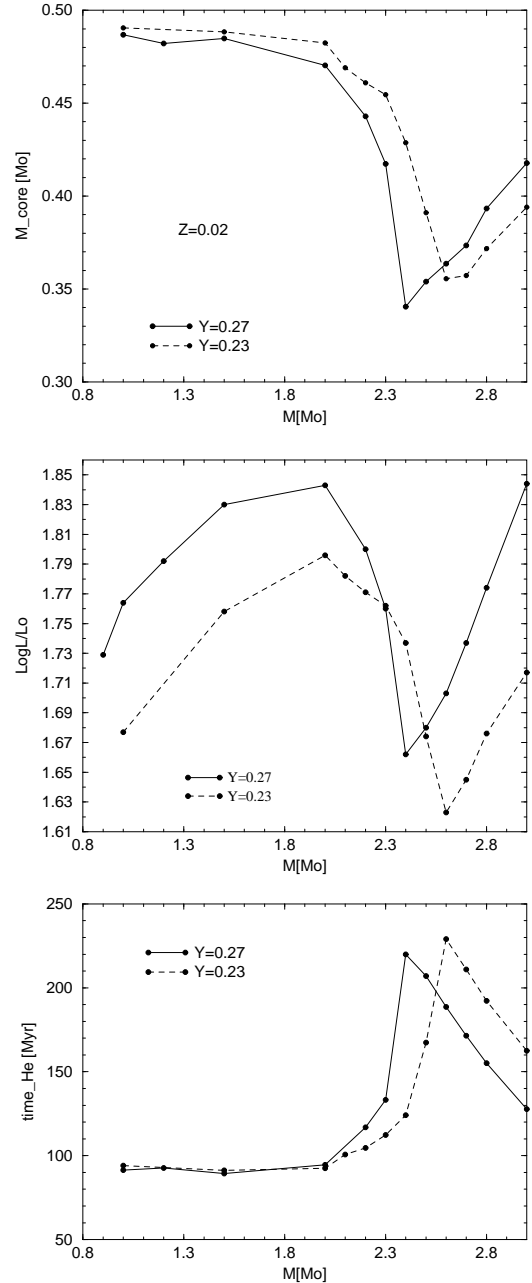}
\caption[]{Selected quantities depicting the behavior of stellar models 
across the Red Giant Transition as computed for solar metallicity (Z=0.02) 
and for the two choices about the amount of original helium Y=0.27 or Y=0.23.
Top to bottom: i) the mass of the He core at the beginning of the major
phase of central He burning, ii)the predicted luminosity of the same models
and, iii)  the predicted He burning lifetime.}
\label{McLogL}
\end{figure}

Fig. 1 gives selected quantities concerning the behavior of He burning
models 
across the RGB-pt as computed for Z=0.02 and the two
alternative assumptions Y=0.27 or 0.23. All models have been computed 
according to the theoretical scenario
 already presented  in C99, which incorporates all the
most recent evaluations of the input physics. 
As everywhere in the following,  quantities given in Fig.1 
refer to the first model which, after igniting central He, has already 
reached the HR diagram location where it will spend the major phase of 
central He burning. Let us here notice that the luminosity of these He
burning
models  is not related only to the mass of the He core. Initially
this luminosity increases in spite of the decreasing He core, since
the increased efficiency of the H burning shell overcomes the
decreased output of He-burning reactions.  However, eventually the
decrease of the He core dominates and the luminosity reaches its
minimum, whereas the lifetime in
the central helium burning phase increases following the decrease of
the efficiency of the He burning reactions.  The subject of the RGB-pt has 
been already widely debated in the literature (see, e.g., Renzini \& Buzzoni 1986, Corsi et al. 1994,
Girardi \& Bertelli 1998 and references therein) and here it does not deserve
further comments.

However,  Fig. 2 compares the luminosities given in
Fig.1 with similar results but by
Girardi et al. (1998: G98 hereinafter). To our surprise, one finds that G98
luminosities appear systematically fainter than in C99 by about
$\Delta$log L/L$_{\odot}$ $\sim$ 0.1 $\div$ 0.15, which is far from being a
negligible amount and, in turn, it appears of the right amount to solve the
already quoted M67 discrepancy.

\begin{figure}[h]
\includegraphics[scale=0.50,draft=false]{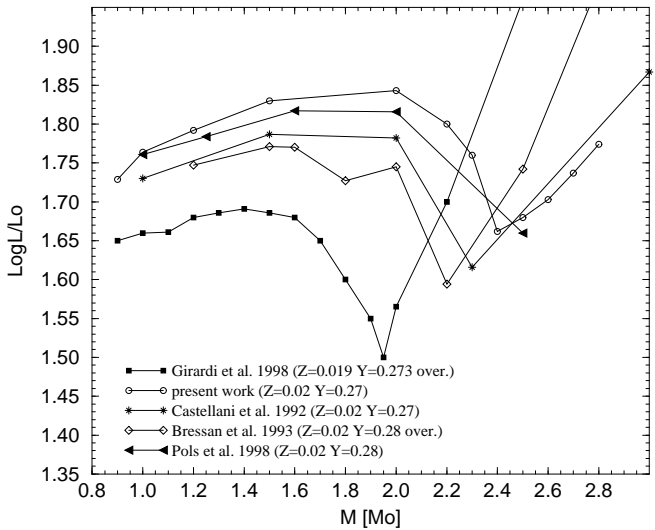}
\caption[h]{He clump luminosity as a function of the stellar mass 
 for Z=0.02 and Y$\approx$0.27 by various authors, as labeled;
``over'' marks evolutions with core overshooting.}
\label{girardi}
\end{figure}

Such an evidence prompted us to investigate similar data in the 
literature, aiming 
to find the origin of such a difference.  To this purpose,  
the same Fig. \ref{girardi}
shows the predictions already given in the literature on the basis of
of the Frascati or Padua evolutionary codes before the last updating of
the input physics.  The figure gives the comforting
evidence that luminosities from Castellani et al. (1992)
appear in rather good agreement with similar data by Bressan et al. (1993).
As we will further discuss in  the next section,
the slight  underluminosity and the little difference in the RGB-pt mass
of Bressan et al. models is only the
expected consequence of their adoption of a moderate core 
overshooting scenario.

The same figure shows that the updated input physics adopted
in  C99 has the effect of
increasing the luminosity of the models with a degenerate progenitors,
according to the discussion already given in Cassisi et al.
(1998) and in reasonable agreement with stellar models recently
presented by Pols et al. (1998). However, one also finds 
that the new input physics in the Padua models
(Girardi \& Bertelli 1998, Girardi et al. 1999) has the opposite effect, 
sensitively decreasing the predicted luminosities. 
As a whole, one finds that uncertainties on  predicted luminosities
can be even larger than $\Delta$logL$\sim$0.1, leaving
an unpalatable uncertainty in the current evolutionary scenario.

On general grounds, one expects that the quoted differences are the
results of differences either in the input physics or in the assumptions
about the efficiency of macroscopic mechanisms, like core overshooting, 
which can affect the evolution of stellar structures. 
To discuss this point, in the next sections we will investigate the range
of variability in current theoretical predictions,  as produced by
the various assumptions governing the evolutionary behaviour.

\section{Theoretical uncertainties in predicted luminosity.}

Making reference to the set  of models presented in 
C99, in this section we will explore the influence 
on central He burning models of several assumptions 
concerning these structures, namely,
i) the efficiency of core overshooting mechanisms,
and,  ii) the effect of mass loss.
In this way we aim to reach a clear insight on the "solidity"
of the result one is dealing with in the literature.

\begin{figure}[h]
\vspace{-5cm}
\hspace{-4cm}
\includegraphics[scale=0.80,draft=false]{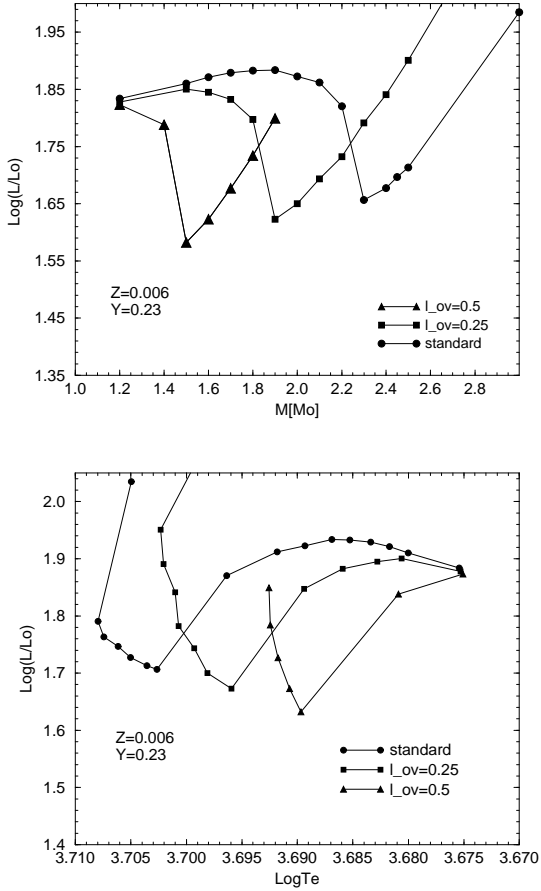}
\vspace{-6cm}
\caption[]{Upper panel: The luminosity of He burning models
with Z=0.006 and Y=0.23 as a function of the stellar mass and for 
the various labeled assumptions about the efficiency of core overshooting
(l$_{-}$ov), see text. Lower panel: the run of the same models but in the
HR diagram.
              }
         \label{over}
   \end{figure}

Fig. \ref{over} (upper panel) shows the effect 
on the model luminosity of selected choices about the efficiency 
of core overshooting 
when the original stellar mass  is varied between 1 and 3 M$_{\odot}$,
while 
the lower panel in the same figure adopts the G98 
representation to show the run of the same models in the HR diagram. 
Labels in these figure give the adopted amount of extramixing (in unity of
the
local pressure scale height) around the convective cores.
In passing, note that comparison between this figure and Fig.1
gives the already known evidence that the RGB phase transition 
shifts to lower masses as the metallicity decreases.

As already known, one finds that overshooting decreases the mass of
the RGB-pt (although it then occurs at  a larger age) 
and, correspondingly, that the maximum luminosity reached by 
the models before the transition decreases. 
However, one finds that for moderate amounts of overshooting
such a decrease is rather small and, in any case, models with 
masses of the order of 1.2 M$_{\odot}$ or lower are little affected by such
a
mechanism. In addition the minimum luminosity attained at the RGB-pt
varies by only $\Delta$logL/L$_{\odot} \approx 0.03$ between a standard 
model and a model with $l_{ov}$=0.25. 
 

Thus the differences in the assumptions about the efficiency of
overshooting
can hardly be at the origin of the differences in Fig. \ref{girardi}  
and, in turn, they cannot be used to reconcile Pols et al. (1998) or C99
computations with M67 or Hipparcos constraints.

  \begin{figure}[h]
\includegraphics[scale=0.50,draft=false]{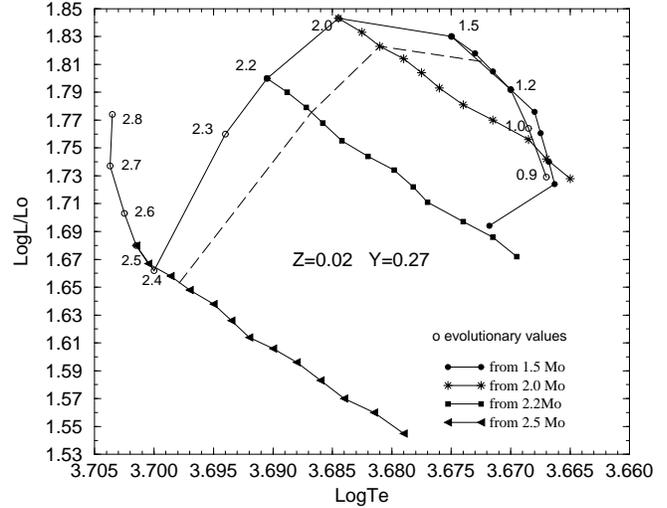}
   \caption[]{The HR diagram location of  He
   burning structures under different assumptions about the amount
  of mass loss in the pre He burning phase. Open circles give
the location of models without mass loss with the labeled values of
the stellar masses. The shift in the HR diagram expected by
decreasing the total mass by step of 0.1 M$_{\odot}$ is shown
for the four labeled values of the original mass. Dashed line
gives the expected distribution for models having lost 10\%
of their original mass.
}
   \label{massloss}
   \end{figure}

The effect of mass loss deserves a bit more discussion. Here we will assume
that mass loss occurs in the advanced phase of H shell burning, so that
the internal structure of the He burning star is not affected by such an
occurrence, which  only decreases the amount of envelope surrounding 
the central He core. Under this assumption, the effect of mass loss
on He burning models can be easily computed by simply decreasing the 
envelope of the constant-mass model.  Fig. \ref{massloss} maps
the effect in the HR diagram of different amount of 
mass loss from the selected models. The behavior depicted by data
in this figure can be easily understood as follows:

i) As long as models develop  strong electron degeneracy (i.e., 
 for masses lower or of the order of 1.5 M$_{\odot}$) the mass
of the He core at the He ignition is the result of RGB evolution.
As a consequence it is largely independent of the evolving mass
and, in addition, it is little affected by mass loss
(see, e.g., the discussion in Castellani \& Castellani 1993),
thus He burning  models with mass losing progenitors behave like models
without mass loss but starting their evolution with the actual mass of
the He burning   model. In this case the evolutionary
behavior with mass loss can be easily predicted in terms of 
canonical models without mass loss. 

ii) For larger stellar masses the He core at the He ignition is largely 
connected to the extension of the  convective cores in the previous 
MS structures. As a consequence, the mass of the He core in He
burning structures is now a sensitive function of the original 
stellar mass, whereas it keeps being little affected by  mass loss,  
which mainly occurs in the post MS phases. In such case theoretical 
expectations for the luminosity abandon the location of canonical models, as shown in 
Fig. 4. 

As an example, Fig. \ref{massloss} gives
theoretical predictions assuming  for all stars a common mass loss of
10\%.

 This is intended to be a useful illustration of the effect we get 
assuming reasonable mass loss along the RGB. However, one should keep 
in mind that the real situation may well be more complicate. Assuming 
the Reimers (1975) mass-loss rates, for instance, one finds that stars 
of higher mass lose less mass along the RGB. This is so
because they have lower RGB-tip luminosities, evolve at higher effective 
temperatures, and have a shorter RGB lifetime. In this case,
stars more massive than about 1.5~$M_\odot$ would lose negligible
amounts of mass on the RGB, contrarily to the lowest-mass ones
(see fig.~6 in Girardi 1999). 
In the context of the present discussion, this 
means that mass-loss is expected to be less effective exactly
in the mass range in which it would more affect the luminosity of 
the He-burning models, i.e. in the left part of Fig.~4.

\section{The Hipparcos constraints}

In Sect. 3 we have explored the influence of physical
effects which are badly understood and the treatment of which is
usually done in parametrized form in stellar evolution codes. 
Having discussed the scenario of these theoretical uncertainties, let us
here attempt to use Hipparcos constraints to test the already
quoted different predictions concerning the luminosities of He burning
stars. Following the careful discussion given by G98
about the clump population, one finds that for any reasonable
assumption about the age spread of field stars one predicts the bulk
of the clump to be populated by stars below the transition mass,
whose luminosity is practically insensitive to assumptions about
the efficiency of overshooting. 
Coming back to Fig.2, one finds that at a solar metallicity 
C99 predictions give for these stars a luminosity larger than G98 by 
about $\Delta$logL=0.1. More in general, comparison of data in G98 and C99
shows that the two evolutionary scenarios have a rather similar
dependence of luminosities on the chemical composition, thus 
with the quoted systematic difference at any given metallicity.

"Sic stantibus rebus", the already quoted evidence that G98 evolutionary 
scenario appears able to nicely fit the Hipparcos mean magnitude of
clumping He burning stars, implies that C99 must predict too luminous
giants, running against the Hipparcos evidence. This has been
confirmed by independent simulation of the clump population
based on C99 evolutionary tracks, as transferred into the CM diagram 
by adopting  model atmospheres by Castelli et al.(1997a,b). 
Data in Fig.1 gives the additional evidence that reasonable 
variations in the assumed original He content cannot decrease the
C99 predicted luminosity by the required amount. Therefore
one concludes that G98 evolutionary scenario works better.
   
However, for the sake of the discussion one has to notice that 
there is -at least in principle - a way to reconcile  C99 prediction 
with Hipparcos observations.  Fig. 4
in this paper shows that a substantial amount of mass loss could lower
the C99 predictions by the required amount of about 
$\Delta$logL$\sim$0.1. As an example, one would require
a mass loss by about 0.9 M$_{\odot}$ for a 2.0  M$_{\odot}$, and
by about 0.6  M$_{\odot}$ for a 1.5  M$_{\odot}$ model.
This, however,  appear  a too large requirement vis-a-vis
current estimates for mass loss. Taking also into account the uncertainties
 on  evolutionary parameters 
of the field population in the solar neighborhood, we regard the
previous discussion not as a proof, but at least as a 
suggestion that the most updated models, as C99 are, when dealing
with the progeny of degenerated RG tend
to give too luminous He burning models.

\section{Model differences}

It appears  of obvious relevance to 
address the problem of the discrepancies between the models by
G98 and C99, trying to understand them in terms of 
different descriptions of the input physics. We leave out differences
in the actual treatment of the input physics (e.g.\ interpolation in
and between opacity tables), which we think are of smaller influence
and might contribute to  a minor part of the variations in the clump luminosities
as shown in Fig.~2. 

\begin{figure}[h]
\includegraphics[scale=0.50,draft=false]{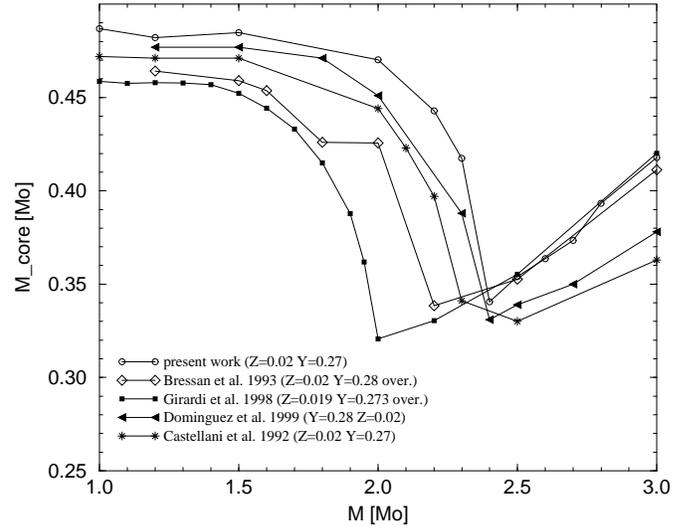}
\caption[]{Comparison between the helium core mass at helium ignition
from G98 and C99 models. Chemical composition as labeled.}
\protect\label{f:mccomp}
\end{figure}

Since the most important quantity determining the core helium
burning luminosity is the core mass at the helium flash, we will
concentrate on this parameter. Fig.~\ref{f:mccomp} shows that the 
1 M$_{\odot}$ G98 models
have core masses lower by $\approx 0.03\,M_\odot$ compared to C99 at
the time of helium ignition at the RGB tip. Fig.~\ref{f:mccomp} 
is made for solar composition but we checked that even for other
metallicities and helium abundances the differences in the He core mass are very similar.

We have identified the following differences in the input physics used
in the two evolutionary programs (``Padua'' for G98 and
``FRANEC'' for C99 models) under consideration
(other aspects such as reaction rates, electron screening,
mixing-length formalism, etc., are to great extent identical): 

\begin{enumerate}	
\item {\em Plasma neutrino emission}: G98 use Munakata et al.\ (1985) for $\log T
> 7.7$, while C99 use Haft et al.\ (1994);
\item {\em Electron conduction}: G98 employ Hubbard \& Lampe (1969), C99
Itoh et al.\ (1983);
\item {\em Radiative opacities}: the '92 tables by OPAL (Rogers \& Iglesias
1992) are used in G98, while the '96 ones (Iglesias \& Rogers 1996)
are those in C99;
\item {\em Equation of state}: C99 use the new OPAL-EOS tables (Rogers et al.\
1996) extended by that of Straniero (1988) in those regions where the
OPAL-EOS does not exist (see Cassisi et al. 1998 for more
details). G98 use their own analytic EOS, which takes into account
partial ionization of hydrogen and helium, degeneracy and Coulomb-effects, 
and which has been described in Girardi et al. (1996).
\end{enumerate}
The influence of some of these differences could be investigated quite
easily because the two codes  to some extent allow
the selection of several sources of input physics. The tests were done
for different cases of initial chemical composition and mass.

\paragraph{Neutrino emission.}
For the composition $Y=0.27$, $Z=0.02$ and an initial mass of
$1.2\,M_\odot$ we find that $M_c$ decreases from 0.482 to 0.476 (-0.006)
$M_\odot$, if we switch from the Haft et al.\ (1994) back to the older
Munakata et al.\ (1985) neutrino emission rates (FRANEC code). This is
the same result as Cassisi et al.\ (1998; models 8 and 7 in their
Table~1) obtained for an $0.8\,M_\odot$ model with $Y=0.23$, $Z=0.0001$
and also identical to what we find in the case of the Padua-code
for the same model. In the latter case, we also verified that
using the Munakata et al.\ (1985) emission down to $\log T = 7.4$
increases the He core mass by $\approx 0.001\,M_\odot$. Therefore,
the total budget of core mass reduction due to the G98 treatment of
neutrino emission physics amounts to $0.007\,M_\odot$. 
Recall that for the $1.2\,M_\odot$ model, the G98 value for $M_c$ is
$0.024\,M_\odot$ smaller than that of C99.

\paragraph{Opacities.} 
We checked the effect of electron conduction opacity by
switching in the FRANEC code from Hubbard \& Lampe (1969)
to Itoh et al.\ (1983) electron conduction opacities.
The radiative opacities in these test cases are of a generation older
than OPAL, but the differential effect of changing condution opacities
can safely be assumed to be largely independent of the radiative
opacities. The test model was a $1.5\,M_\odot$ star of $Y=0.27$,
$Z=0.02$. The use of the older conductive opacities (i.e.\ those used
by G98) leads to a reduction of $0.008\,M_\odot$ in $M_c$.

The influence of switching from the '92 to the '96 OPAL opacities was
not tested, but according to our experience it should be minor
compared to that of the electron conduction.

\begin{figure}[h]
\includegraphics[scale=0.50,draft=false]{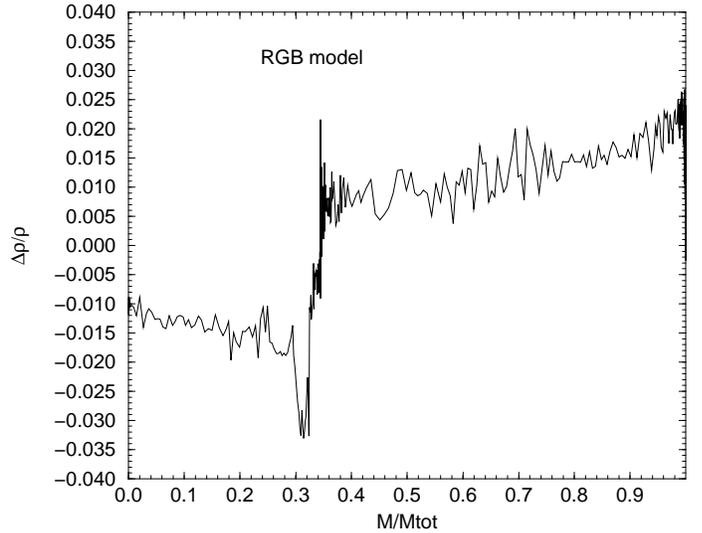}
\caption[]{ Relative difference in density stratification between
C99 and G98 EOS for an RGB model (see text).} 
\protect\label{f:eoscomp}
\end{figure}

\paragraph{Equation of state.}
The EOS being a complicated part of both programs, we could not easily
exchange one for the other. However, we could perform the following
test: we took the pressure and temperature stratification of a C99-model
on the RGB and applied the G99-EOS to it in order to obtain the
density. We compared with the original C99 density
stratification. The result is shown in Fig.~\ref{f:eoscomp} for a RGB
model with  Y=0.238, Z=0.004: while the core would have densities
higher by about 1.5\% in the Padua code, the envelope would be less
dense by 1--2\%. Both effects reflect the evolutionary changes along
the RGB, such that a G98-model would appear to be more evolved than a C99
one. Therefore, also the difference in the EOS is expected to lead to
a lower core mass at helium ignition for the G98 calculations.

From these three investigations we conclude that with the differences
in neutrino emission rates and conductive opacities we can explain
almost 60\% of the discrepancy in core mass. This moves the G98 core
masses already within the general spread of results. At least part of
the remaining difference can be ascribed to the EOS. 

Finally, we have compared results with three further codes, which use
almost identical input physics as C99, in particular with regard to
radiative opacities, neutrino emission and electron conduction. With regard to the
Garching stellar evolution code (see, e.g., Schlattl \& Weiss 1999) we
find that $M_c$ is $0.004\,M_\odot$ larger in the C99 models for a
$0.8\,M_\odot$ star of $Y=0.23$ and two metallicities,
$Z=0.0001,\> 0.001$. This small difference can be understood
as being a consequence of the fact that more ($0.004$) helium is
dredged up in the C99 models. 

\begin{figure}[h]
\vspace{-4cm}
\hspace{-4cm}
\includegraphics[scale=0.80,draft=false]{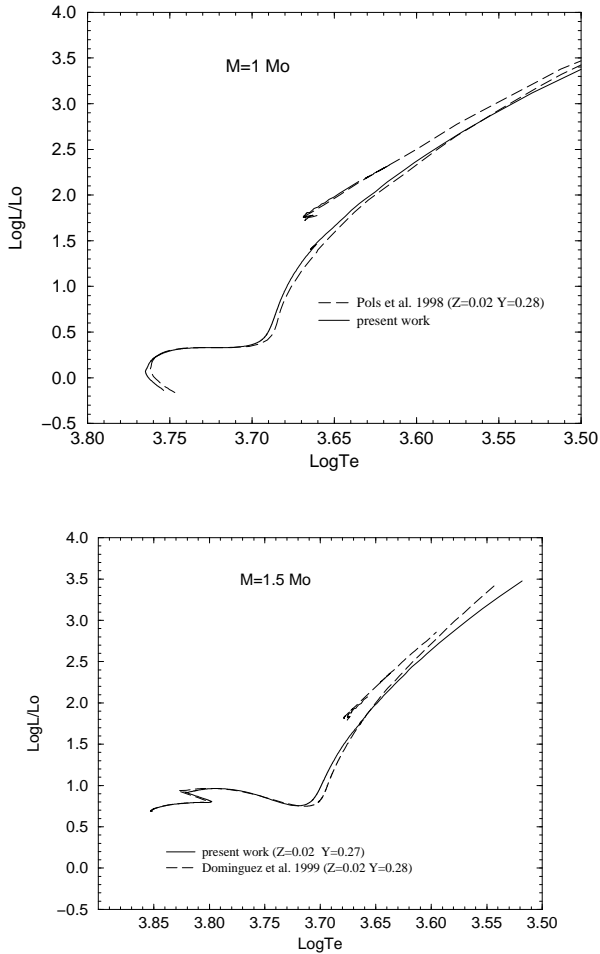}
\vspace{-5cm}
\caption[]{Upper panel: comparison between the 1 M$_\odot$ model with solar
compositions
from Pols et al. (1998) and from C99; lower panel:
 comparison between the 1.5 M$_\odot$ model with solar compositions
from Dominguez et al. 1999 and from C99.}
\protect\label{f:spscomp}
\end{figure}

Finally, evolutionary calculations by Pols et al. (1998) and
Dominguez et al. (1999) with their
respective codes (Pols et al. 1998; Straniero et al. 1997)
reveal for various chemical compositions a very high degree of agreement
with those of C99 as shown in Fig~\ref{f:spscomp},
including $M_c$, which differs by less than
0.01 M$_\odot$ between Dominguez and C99.

We therefore can conclude that more than half of the difference in the
core mass at helium ignition between G98 and C99 can be removed by
adopting the same neutrino emission rates and electron conduction
opacities; using identical EOS would lead to further convergence,
although we cannot quantify this point. At the same time, codes with
identical physics do indeed result in very similar core masses which
differ at a level of a few $10^{-3}\,M_\odot$ only. Therefore, the
differences between the G98 and C99 results concerning this quantity
are well understood as a consequence of different physical inputs.
Note that taking from the literature (see e.g. Sweigart \& Gross 1978) 
$\Delta$logL/L$_{\odot}$ $\approx 3.4~\Delta M_c$ for stars with
degenerate progenitors, {\bf the difference in $M_c$ explains the
difference in luminosity among the various models in Fig.2, with the exception
of predictions by Bressan et al. (1993) that reveal the contribution of
some other difference than the core mass only.}
As a conclusion, the different predicted luminosities we are dealing
with appear as the natural results of evolutionary codes with
different -- but in both cases reasonable -- input physics and thus
as an example of the intrinsic unavoidable uncertainties in any
current evolutionary scenario.

\section{Conclusions}

As it is well known, and as recently explicitly
reiterated, e.g., in Castellani (1999), the widespread use
of the Henyey algorithm in producing stellar models assures
that different codes with identical input physics must produce
quite similar results. In this paper we presented several
evidences, as taken from the literature, supporting such
an occurrence. Thus evolutionary models are as "good"
as the adopted input physics is. 
In this paper we have however shown that models based on different, 
but all "reasonable", physics inputs may have sizable differences
in the predicted luminosity of He burning low mass stars.

Testing these theoretical predictions on the absolute
magnitudes given by the Hipparcos satellite for He burning 
stars clumping in the field, one finds several indications
for preferring the luminosity predicted by the Padua code,
as used in  Girardi et al. (1998). Even if the argument is
still open to further investigation, we feel that the 
discussion presented
in this paper raises - at least - serious doubts on 
the capability of the most recent "updated input physics"
of correctly predicting the luminosity of He burning stars as originated
from progenitors which underwent strong electron degeneracy. 
Therefore giving a serious warning about the uncritical
acceptance of theoretical predictions concerning the luminosity
of HB stars in galactic globulars and, in turn, all the
related quantities. 

In this respect, we note that observational evidences concerning
the luminosity of globular HB stars are far from being clear.
Caputo et al.  (1999) have  recently discussed RR Lyrae variables 
in the globular M5 showing that their pulsational properties 
possibly suggest a luminosity lower than predicted by C99 models.
However, Cassisi et al. (1999) and Salaris \& Weiss (1998) used
HB stars brighter than G98 predictions to derive cluster distance moduli
which appear in excellent agreement with some independent
evaluations based on Hipparcos parallaxes for field subdwarfs.
Here we can only conclude that the matter deserves further
investigation, to decide whether or not the mass loss can
reconcile "most updated" stellar models   with
Hipparcos constraints for the clump of He burning stars 
in the solar neighborhood. 

Before closing the paper, let us notice that all the computations
discussed in the paper  agree in
predicting  He burning stars systematically brighter when the metallicity
is decreased.  The theoretical predictions are illustrated, e.g., in the
figure 1 of G98, and figure 12 of C99.
This has important implications for the method of distance
determination based on clump stars, since it indicates
that the clump population observed in nearby galaxies may have an intrinsic
luminosity different from the local stars sampled by Hipparcos. In the
case of the Magellanic Clouds, this difference may amount to 
0.2 - 0.3 mag, as inferred by  Cole (1998) and G98. 

\section{Acknowledgments}

It is a pleasure to thank O.R. Pols and O. Straniero for
kindly providing us with the evolutionary tracks from Pols et al.(1998)
or Dominguez et al. (1999). A. Chieffi and M. Limongi are acknowledged
for helpful discussions. The finantial support of the ``Ministero della
Universit\'a e della Ricerca Scientifica e Tecnologica'' to the project
{\em Stellar Evolution} is kindly acknowledged.

\end{document}